\documentclass[conference]{IEEEtran}
\IEEEoverridecommandlockouts

\usepackage{cite}
\usepackage{amsmath,amssymb,amsfonts}
\usepackage{algorithmic}
\usepackage{graphicx}
\usepackage{hyperref}
\usepackage{textcomp}
\usepackage{xcolor}
\def\BibTeX{{\rm B\kern-.05em{\sc i\kern-.025em b}\kern-.08em
    T\kern-.1667em\lower.7ex\hbox{E}\kern-.125emX}}

\makeatletter

\def\ps@IEEEtitlepagestyle{
    \def\@oddfoot{\mycopyrightnotice}
    \def\@evenfoot{}
}
\def\mycopyrightnotice{
}



\makeatletter
\newcommand*\titleheader[1]{\gdef\@titleheader{#1}}
\AtBeginDocument{%
  \let\st@red@title\@title%
  \def\@title{%
    \bgroup\normalfont\small\raggedright\@titleheader\par\egroup
    \vskip0.1em\st@red@title}
} \makeatother

\title{Design and Analysis of a Grid-connected DC Fast
Charging Station for Dhaka-Chittagong Highway\\
}

\titleheader{}
    
\begin{document}

\author{\IEEEauthorblockN{Alif Ahmed, Minhajur Rahman$^{*}$, Mohammad Jawad Chowdhury, Khandakar Abdulla Al Mamun}
\IEEEauthorblockA{\textit{Dept. of Electrical \& Electronic Engineering} \\
\textit{Int'l Islamic University Chittagong}\\
Chittagong, Bangladesh \\
alif.ahmed.eee@gmail.com, $^{*}$fahad061299@gmail.com, jawad.iiuc.eee@gmail.com, k.a.a.mamun@gmail.com}
}

\maketitle

\begin{abstract}
The growing adoption of electric vehicles (EVs) necessitates the development of efficient and reliable charging infrastructure, particularly fast charging stations (FCS) for addressing challenges such as range anxiety and long charging times. This paper presents the design and feasibility analysis of a grid-connected DC fast charging station for the Dhaka–Chittagong highway, a critical transportation corridor in Bangladesh. The proposed system incorporates advanced components, including a step-down transformer, Vienna Rectifier, and LC filter, to convert high-voltage AC power from the grid into a stable DC output. Simulated using MATLAB Simulink, the model delivers a peak output of 400V DC and 120 kW power, enabling rapid and efficient EV charging. The study also evaluates the system's performance, analyzing charging times, energy consumption, and distance ranges for representative EVs. By addressing key technical, environmental, and economic considerations, this paper provides a comprehensive roadmap for deploying fast charging infrastructure, fostering EV adoption, and advancing sustainable transportation in Bangladesh.
\end{abstract}

\begin{IEEEkeywords}
electric vehicle, fast charging station, converter.
\end{IEEEkeywords}

\section{Introduction}
Electric vehicles (EVs) are becoming essential in the global effort to reduce greenhouse gas (GHG) emissions and tackle the environmental challenges posed by the transportation sector \cite{chapman2007transport}. Transportation sector accounts for 24\% of global carbon emissions. \cite{mishra2021comprehensive}. This alarming statistic has created a worldwide shift from fossil fuels to cleaner, more sustainable energy alternatives. EVs, with their zero emissions, increased efficiency, and long-term cost-effectiveness, offer a promising solution to the dual problems of environmental degradation and rising energy demands. They not only address urgent environmental concerns but also represent a transformative opportunity to reshape the global transportation landscape.

\begin{figure*}[t]
\centerline{\includegraphics[width=1\linewidth]{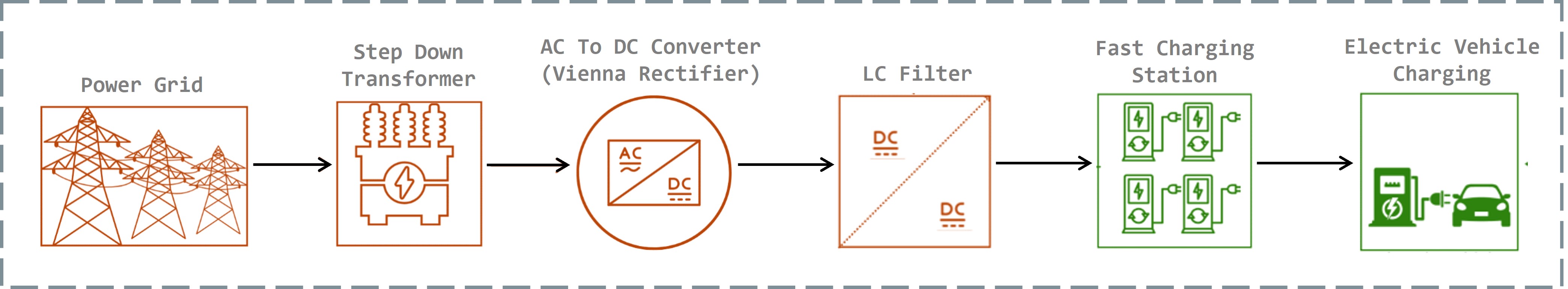}}
\caption{Charging process of an electric vehicle from the power grid using a fast charging station. High-voltage AC power from a three-phase grid is stepped down via a transformer and converted into DC power using a Vienna Rectifier. An LC filter is incorporated to remove noise and stabilize the DC output, ensuring a clean and reliable power supply for efficient and rapid vehicle charging.}
\label{fig:pipeline}
\end{figure*}
Despite their potential, the adoption of EVs faces significant challenges, particularly in developing robust charging infrastructure. One of the most common obstacles is range anxiety-the fear of depleting an EV's battery before reaching a charging station \cite{gupta2024charging}. This issue is compounded by the lengthy charging times of conventional slow chargers, which typically take 6–8 hours for a full charge. Fast charging stations (FCS) present a viable solution, dramatically reducing charging times to as little as 15–60 minutes \cite{botsford2009fast}. Operating at high power levels of 50 kW to 240 kW and capable of delivering currents up to 400 A, these stations provide the rapid energy replenishment required for EVs \cite{schwarzer2015current}. The average capacity of standard EV batteries is around 60.1 kWh, supporting an average mileage of 317 km \cite{wang2021grid}. This makes them especially crucial for long-distance travel, high-traffic routes, and urban centers. However, the development of FCS infrastructure involves significant challenges, including the need for advanced technical systems, substantial energy supplies, and the integration of renewable energy sources to ensure sustainability.

In Bangladesh, the transportation sector heavily relies on fossil fuels, creating a dual challenge of environmental harm and energy inefficiency. As of 2022, over 5.3 million internal combustion engine (ICE) vehicles were registered, contributing significantly to the country’s GHG emissions, which accounted for 14.5\% of national carbon emissions in 2019 \cite{islam2021renewable}. Although hybrid electric vehicles (HEVs) and light electric vehicles (LEVs) are gradually gaining popularity, the adoption of more advanced EVs, such as battery electric vehicles (BEVs) and plug-in hybrid electric vehicles (PHEVs), remains limited. This is largely due to the absence of comprehensive EV policies, supportive incentives, and sufficient charging infrastructure. LEVs, used primarily for short-distance commuting since 2009, consume approximately 500 MW of electricity daily, further straining the national power grid \cite{karmaker2018feasibility}.

To facilitate the transition to sustainable transportation in Bangladesh, fast charging stations are crucial. Placing FCS along high-traffic routes, such as the Dhaka-Chittagong highway, can overcome key barriers to EV adoption. These stations would reduce range anxiety, promote EV use, and help lower the country's reliance on fossil fuels and GHG emissions. Additionally, integrating renewable energy sources like solar and wind into FCS infrastructure can improve sustainability, reduce environmental impacts, and ease the strain on the national power grid.

This paper explores the design and analysis of fast charging stations for EVs, focusing on their technical, environmental, and economic aspects. It examines how FCS can support the growth of EVs, reduce the nation’s carbon footprint, and promote cleaner transportation systems. By addressing the challenges and opportunities associated with developing FCS infrastructure in Bangladesh, this study aims to provide a roadmap for accelerating the adoption of EVs and fostering a more sustainable future for transportation.

\section{Criteria and Requirements for Fast Charging of EVs}

Fast charging of EVs requires meeting several criteria and technical specifications to ensure compatibility, efficiency, and safety. The charging levels are categorized into three types based on power and speed:
\begin{enumerate}
    \item \textbf{Level 1 Charging (120V):} This basic level uses standard electrical outlets, offering the slowest charging speeds.
    \item \textbf{Level 2 Charging (240V):} Faster than Level 1, it provides a quicker option for EV owners.
    \item \textbf{Level 3 Charging (DC Fast Charging):} This category includes rapid chargers delivering power outputs from 15 kW to 350 kW. Typical output voltages range from 200 V to 1000 V, and output currents vary between 50 A and 500 A. DC fast chargers significantly reduce charging times, enabling an 80\% charge within 15–45 minutes, depending on the vehicle's capacity and charger specifications.
\end{enumerate}

Charging infrastructure varies depending on its use case. Public charging stations often feature DC fast chargers or AC charging boxes. These stations are designed to meet diverse customer needs while charging a premium for convenience and faster service. Fast charging technology is integral to the growing EV ecosystem. By providing rapid and efficient energy replenishment, these systems address range anxiety and support the widespread adoption of EVs. Each charging type and standard caters to specific requirements, ensuring a flexible and adaptable charging network for all users.

\section{Feasibility Analysis of FCS location and EVs}
\subsection{Site Location and Characteristics}

Dhaka and Chattogram are two vital metropolitan areas in Bangladesh, connected by the 250-kilometer Dhaka–Chattogram corridor. This route is critical for economic activity, with Dhaka serving as the nation’s commercial hub and Chattogram managing 90\% of its imports and exports through its port. A quarter of Bangladesh’s population lives along this corridor, underscoring its strategic importance. The highway has historically faced congestion, long travel times of up to 10 hours, and growing traffic demands, projected to exceed 119,000 daily vehicles by 2040. Infrastructure improvements, including a four-lane upgrade in 2016 and plans for an access-controlled expressway, aim to address these challenges.
\begin{figure}[t]
\centerline{\includegraphics[width=1\linewidth]{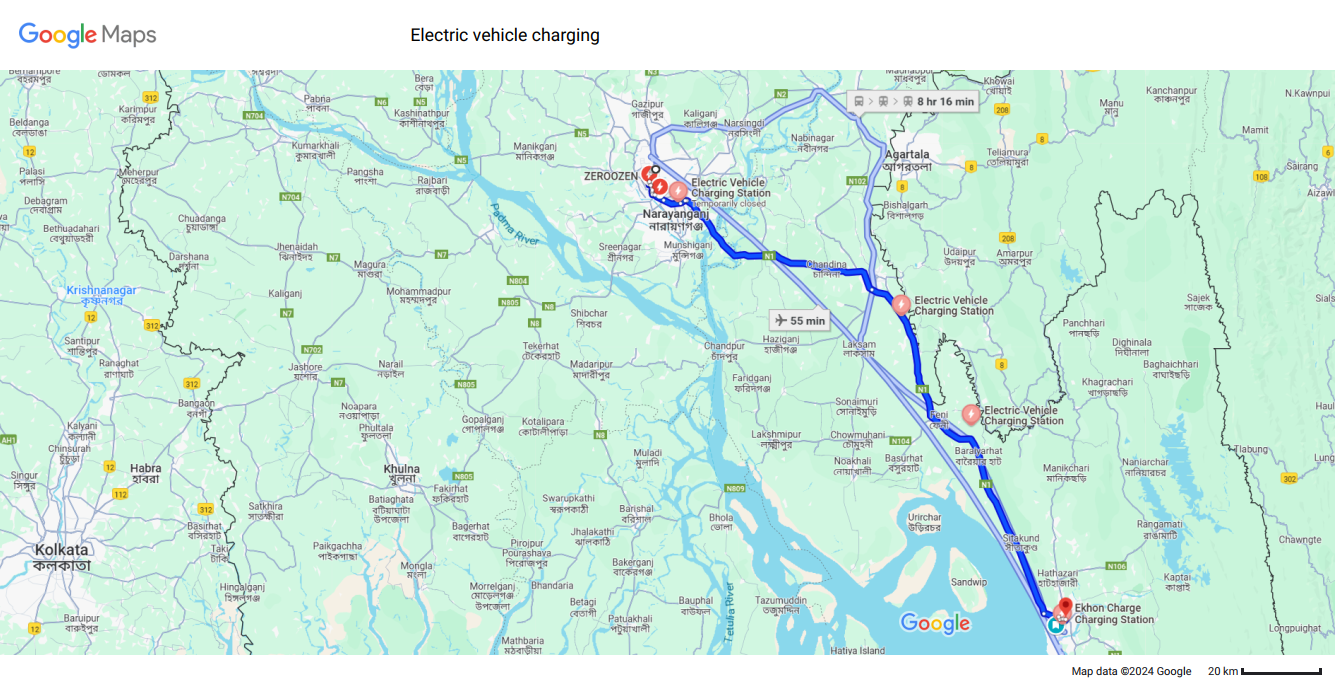}}
\caption{Map of the Dhaka–Chattogram Highway, a critical 250-kilometer transportation corridor connecting Bangladesh's commercial hub, Dhaka, with its primary port city, Chattogram.}
\label{fig:map}
\end{figure}
To further enhance the corridor's efficiency and sustainability, integrating fast charging stations is essential. These stations would support the growing adoption of electric vehicles (EVs) by enabling rapid charging and mitigating range anxiety. Fast charging infrastructure would not only promote EV adoption but also reduce emissions, align with sustainability goals, and improve the overall efficiency of this critical transportation route.

\begin{figure*}[t]
\centerline{\includegraphics[width=1\linewidth]{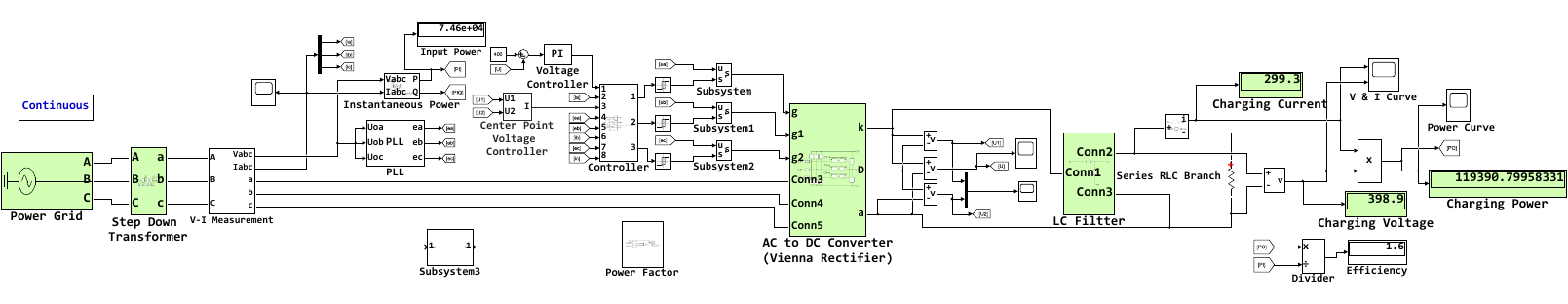}}
\caption{Simulink model of our proposed DC FCS, illustrating the power conversion process from a three-phase AC power grid to a stable DC output. The model includes key components such as the step-down transformer, Vienna Rectifier, and LC filter.}
\label{fig:simulink}
\end{figure*}

\subsection{ Electrical Power and Charging Outlet Voltages of EVs} 
To calculate the number of vehicles that can be charged by the proposed charging station, it is necessary to select specific vehicles representing a range of power requirements. The charging capacity of the station depends on the power consumption of each electric vehicle (EV), as higher-powered vehicles require more energy and longer charging times, reducing the total number of vehicles that can be charged. Conversely, lower-powered EVs consume less energy, allowing the station to charge a greater number of vehicles within the same timeframe. To represent a diverse range of power needs, five EV models have been selected, covering both high and low power requirements. The data for these vehicles, including battery capacity, real energy consumption, nominal voltage, and fast charging parameters, are summarized in Table \ref{tab:vehicles-data}. This selection ensures a balanced approach to estimating the station's capacity while accounting for variability in vehicle power needs.

\begin{table}[]
\caption{VEHICLES DATA FOR CHARGING STATION DESIGN }
\label{tab:vehicles-data}
\resizebox{\linewidth}{!}{%
\begin{tabular}{lcccc}
\hline
\multicolumn{1}{c}{\textbf{\begin{tabular}[c]{@{}c@{}}Electric \\ Vehicle Model\end{tabular}}} &
  \textbf{\begin{tabular}[c]{@{}c@{}}Battery \\ Capacity\end{tabular}} &
  \textbf{\begin{tabular}[c]{@{}c@{}}Power \\ Consumption\\ (Wh/km)\end{tabular}} &
  \textbf{\begin{tabular}[c]{@{}c@{}}Nominal \\ Voltage\end{tabular}} &
  \textbf{\begin{tabular}[c]{@{}c@{}}Power Acceptance \\ Rate \\ (Fast Charging)\end{tabular}} \\ \hline
BMW iX3 \cite{BMWix3}             & 74.0 kWh & 211 Wh/km       & 400 V & 155 kW DC \\
Ford Mustang Mach-E \cite{Ford}  & 91.0 kWh & 207 Wh/km & 400 V & 150 kW DC \\
Tesla Model 3 \cite{Tesla_Model3}       & 57.5 kWh & 144 Wh/km & 400 V & 170 kW DC \\
Tesla Model Y \cite{Tesla_ModelY}      & 75.0 kWh & 185 Wh/km & 357 V & 250 kW DC \\
Volkswagen ID.4 Pro \cite{Volkswagen} & 77.0 kWh & 193 Wh/km & 400 V & 135 kW DC \\ \hline
\end{tabular}}
\end{table}

\section{Methodology}
The design and implementation of the DC fast charging station for electric vehicles (EVs) are based on a structured approach that combines advanced power conversion technologies sinulated using MatLab simulation tools. This section outlines the methodology used to develop the proposed system, highlighting the key components, their roles, and the simulation process.

\subsection{FCS System Design Component}
The core of our proposed system is illustrated through figure \ref{fig:pipeline}, which provides a visual representation of the structure and functionality of the DC fast charging station. The system begins with an 11 kV three-phase AC power input from the grid. This power undergoes several stages of conversion to deliver a clean and stable DC output suitable for EV charging. Each component in the system has a specific role in ensuring efficiency and reliability:
\begin{enumerate} \item Three-Phase Power Grid 
\item Step-Down Transformer
\item Vienna Rectifier
\item LC Filter
\item DC Power Output
\end{enumerate}

\subsection{Proposed Model }

The proposed model of the DC-DC fast charging station has been developed and simulated using MATLAB and its Simulink environment. This simulation platform enabled the creation of a graphical block diagram, allowing for real-time adjustments and performance evaluation.

\textbf{Power Conversion Process:}
The simulation model comprehensively represents the power conversion process, starting from the input of high-voltage AC power to delivering a stable DC output. Our proposed charging station draws 11 kV AC power from a three-phase grid. This high-voltage input ensures that the system has sufficient capacity to meet the power demands of EVs. A step-down transformer reduces the 11 kV AC to 315V AC. This transformation not only lowers the voltage to a manageable level but also increases the current, meeting the station's power requirements for subsequent stages. The 315V AC is converted to DC power through the Vienna Rectifier, which functions as a three-phase boost converter. This rectifier is configured as a three-level, three-switch system that ensures sinusoidal mains current and delivers a controlled DC voltage output of approximately 400V. During the conversion process, noise and harmonics are introduced into the DC output. To address this, an LC filter is strategically placed after the Vienna Rectifier. This filter removes high-frequency noise, stabilizing the output signal and ensuring the delivery of high-quality DC power suitable for EV charging. The final stage of the model provides a DC output with a voltage of approximately 400V and a current of around 300A. These parameters result in a peak power output of 120 kW, enabling the station to charge EVs quickly and efficiently. The Simulink model incorporates advanced monitoring and control mechanisms to ensure safe and optimal operation.

\begin{figure}[t]
\centerline{\includegraphics[width=1\linewidth]{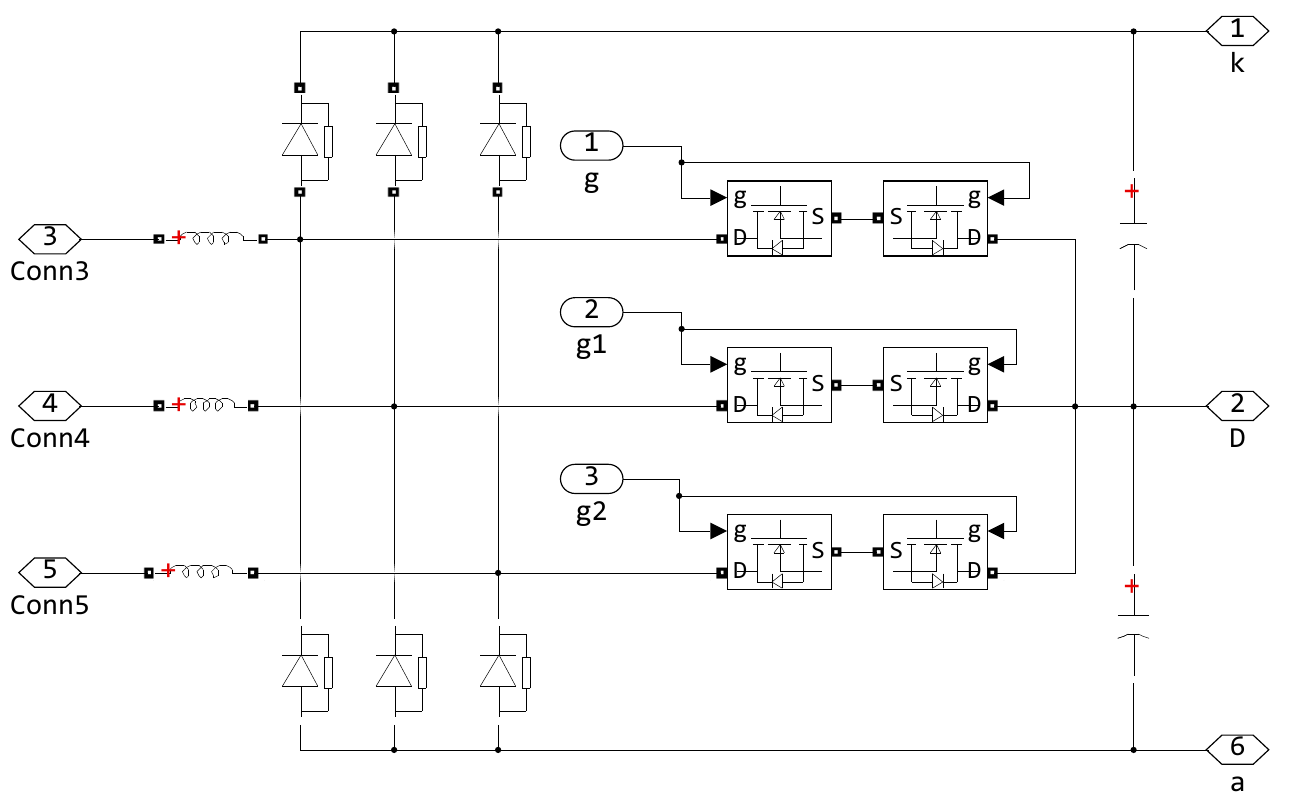}}
\caption{Simulink model of Vienna Rectifier used in our proposed DC FCS. The rectifier, configured as a three-phase boost converter, converts 315V AC power into a nominal 400V DC output. It ensures sinusoidal mains current, controlled DC voltage, and efficient power conversion for electric vehicle charging.}
\label{fig:vienna}
\end{figure}

\textbf{Vienna Rectifier:} The Vienna Rectifier is a critical component in the power conversion process of the DC Fast Charging Station. Its primary function is to convert the 315V alternating current (AC) power, derived from the three-phase grid and stepped down by a transformer, into a stable direct current (DC) output. Configured as a three-phase boost converter, the Vienna Rectifier operates as a three-level, three-switch rectifier. This configuration ensures sinusoidal mains current, which minimizes harmonic distortion and enhances overall system efficiency. By employing advanced pulse-width modulation (PWM) techniques, the rectifier achieves controlled DC voltage output, critical for meeting the precise requirements of electric vehicle (EV) charging systems. The output voltage of the Vienna Rectifier is approximately 400 volts DC, making it suitable for high-power applications like fast charging. Its design enables efficient energy conversion while maintaining stability in the output signal. This stability is vital for ensuring the reliability and safety of the DC Fast Charging Station. The Vienna Rectifier plays a pivotal role in the charging station’s functionality, bridging the gap between the AC power grid and the DC output required for EV charging. Its efficiency and ability to provide a controlled DC output highlight its importance in modern charging infrastructure.

\section{Simulation and Result Analysis}
This section presents the output results from the simulation of the proposed DC Fast Charging Station model using MATLAB Simulink. The focus is on analyzing the output current, voltage, and power over time to validate the performance of the charging station.

\begin{figure}[h]
\centerline{\includegraphics[width=1\linewidth]{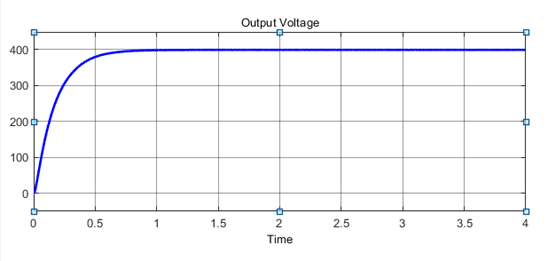}}
\caption{The blue curve illustrates the voltage's behavior, showing an initial surge peaking at 399.1 V, followed by stabilization to ensure a steady and reliable DC output for efficient charging.}
\label{fig:voltage}
\end{figure}

\begin{figure}[h]
\centerline{\includegraphics[width=1\linewidth]{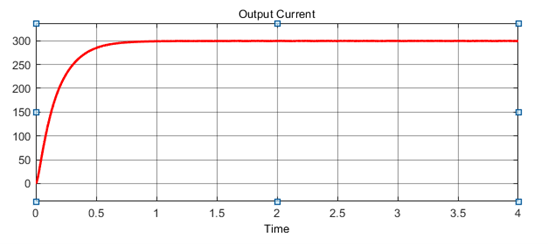}}
\caption{The red curve depicts the current's behavior, showing an initial surge peaking at 299.4 A before stabilizing to a consistent level, ensuring reliable operation during the charging process.}
\label{fig:current}
\end{figure}

Figure \ref{fig:voltage} represents the output voltage values over time in the DC fast charging station. The x-axis of the graph represents time, while the y-axis represents the output voltage. The blue curve on the graph shows the behavior of the output voltage over time. It sharply rises near the beginning, indicating an initial surge in output voltage. This surge reaches a peak of 399.1 volts, as indicated in the box above the graph. After reaching this peak, the output voltage levels off and stabilizes, as shown by the bottom of the curve.

Figure \ref{fig:current} represents the output current values over time in the DC fast charging station. The x-axis of the graph represents time, while the y-axis represents the output current.nThe graph's red curve shows the output current's behavior over time. It sharply rises near the beginning, indicating an initial surge in output current. This surge reaches a peak of 299.4 A, as indicated in the box above the graph. After reaching this peak, the output current levels off and stabilizes, as shown by the plateau of the curve.

\begin{figure}[h]
\centerline{\includegraphics[width=1\linewidth]{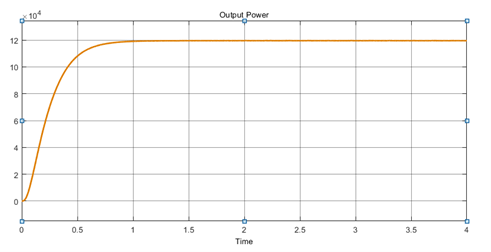}}
\caption{The yellow curve depicts the power's behavior, with an initial surge peaking at approximately 119.5 kW (scaled by $10^4$), followed by stabilization to ensure consistent energy delivery during the charging process.}
\label{fig:power}
\end{figure}
Figure \ref{fig:power} represents the output power values over time in the DC fast charging station. The x-axis of the graph represents time, while the y-axis represents the output power, scaled by 10\^4. The yellow curve on the graph shows the behavior of the output power over time. It sharply rises near the beginning, indicating an initial surge in output power. This surge reaches a peak of approximately 119517.88 power units. After reaching this peak, the output power levels off and stabilizes, as shown by the bottom of the curve.

\subsection{Travel distance analysis}

We analyze the relationship between charging time and travel distance for five electric vehicles, including the Vehicles BMW iX3 , Ford Mustang Mach-E, Tesla Model 3, Tesla Model Y and Volkswagen ID.4 Pro, over charging intervals of 5, 10, 15, 20, 25, and 30 minutes. The findings indicate that charging time increases with distance, though not uniformly, reflecting real-world inefficiencies such as energy losses, battery limitations, and environmental factors. Bar graphs for each vehicle reveal that shorter charging times correspond to limited travel ranges, while longer charging times significantly extend the distances covered. For instance, the Ford Mustang Mach-E and Volkswagen ID.4 Pro demonstrate varying rates of increase in charging efficiency, with distinct distance intervals illustrating the practical implications for charging station infrastructure. Using the BMW iX3 as a benchmark, a 5-minute charge enables approximately 40 km of travel, requiring seven charging stations for a 250 km route, whereas a 10-minute charge extends the range to 80 km, reducing the need to four stations. These findings underscore the importance of strategically optimizing charging station placement to alleviate range anxiety, enhance EV adoption, and enable reliable long-distance travel in Bangladesh.

\section{Conclusion}

\begin{figure}[t]
\centerline{\includegraphics[width=1\linewidth]{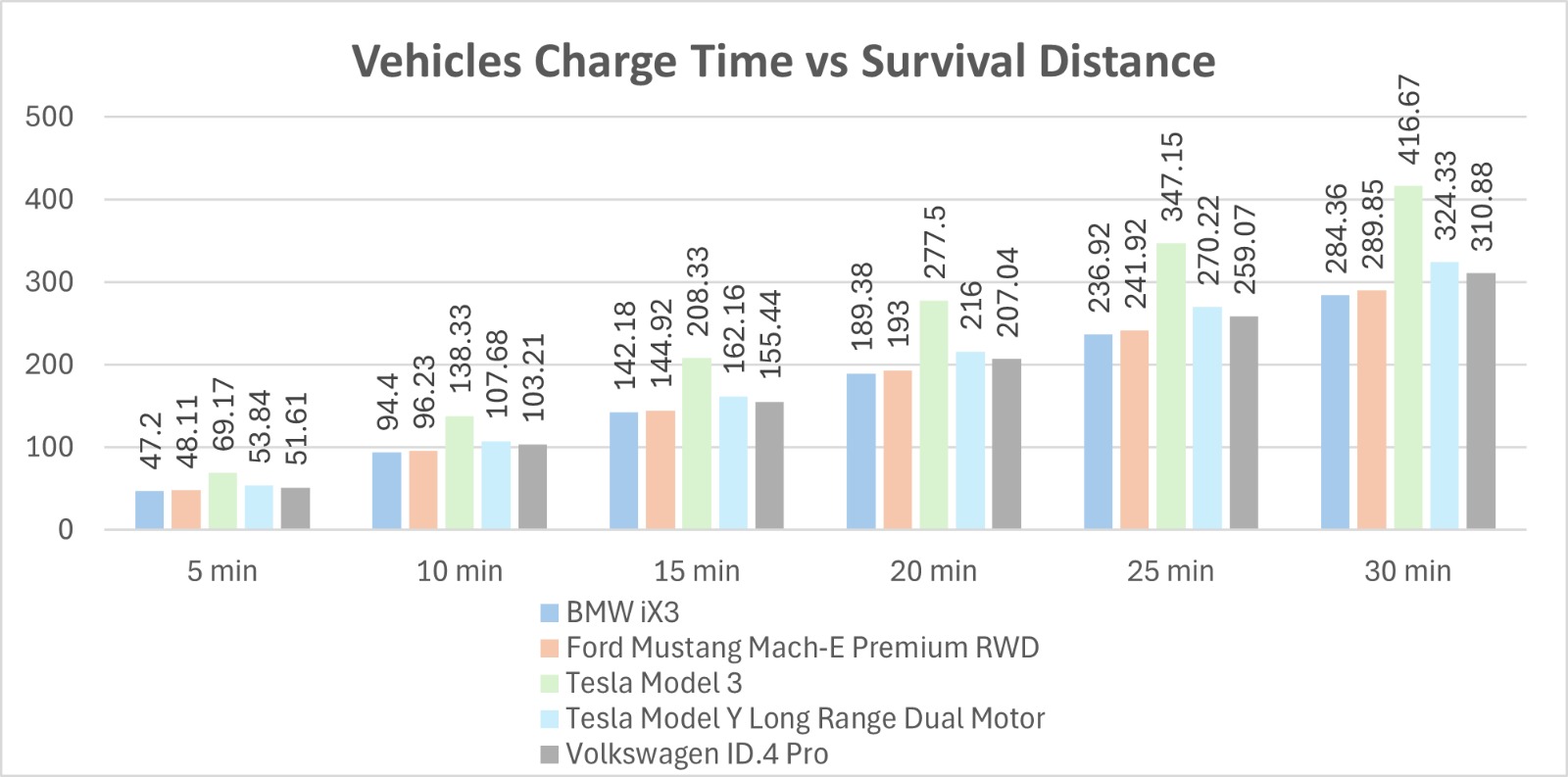}}
\caption{Charging time (minutes) versus EV travel distance (kilometres) analysis for selected electric vehicle models, illustrating variations in efficiency and the impact of real-world factors on charging performance and infrastructure planning.}
\label{fig:analysis}
\end{figure}
This paper analyze the impact of fast charging infrastructure on range anxiety by analyzing the relationship between charging times and travel distances for five popular EV models. Our findings reveal a nonlinear trend influenced by real-world factors such as energy losses and battery limitations, underscoring the need for strategically placed fast-charging stations to ensure seamless long-distance travel. Specifically, our analysis of the 250 km Dhaka-Chittagong highway suggests that four to seven charging stations are required, depending on charging intervals, to mitigate range anxiety effectively. These insights emphasize the critical role of expanding EV infrastructure in promoting sustainable and reliable transportation in Bangladesh. For future work we will focus on optimizing charging station placement, and integrating emerging charging technologies to further enhance EV adoption and address region-specific challenges.

\bibliographystyle{ieeetr}
\bibliography{references}

\end{document}